# Estimating Atmospheric Mass Using Air Density

**L. L. Simpson[1] and D. G. Simpson[2]**

[1]Union Carbide Corporation, South Charleston, West Virginia, USA (retired).

[2]NASA Goddard Space Flight Center, Greenbelt, Maryland USA.

## Abstract

Since the late 19[th] century, several investigators have estimated the mass of the atmosphere. Unlike previous studies, which focus on the average pressures on the earth's surface, this analysis uses the density of air above the earth's surface to predict the mass of the atmosphere. Results are consistent with recent pressure-based estimates. They indicate that changes in the latest estimates can be attributed to improved land elevation measurements between 1 km and 3 km. This work also provides estimates of atmospheric mass by layer and mean and median land elevations.

## Introduction

The mass of air above the earth has long been the subject of scientific investigation and public interest and curiosity. While being a small fraction of the earth's mass, it is a very large value by most standards. It is approximately one fourth of the U.S.A. land mass (land above sea level).

Researchers began reporting estimates of the mass of the atmosphere ($M_A$) in the late nineteenth century. *Verniani* [1965] summarized $M_A$ values referenced from 1892 to 1965. Half of the 14 values are between $5.13 \times 10^{18}$ kg and $5.14 \times 10^{18}$ kg and all of the remaining seven values are slightly larger. Values reported from 1965 are summarized in Table 1.

**Table 1.** Recent estimates of atmospheric mass.

| Author | Year | $M_A$ ($10^{18}$ kg) |
|---|---|---|
| Verniani | 1965 | $5.136 \pm 0.007$ |
| Anderson *et al.* | 1975 | $5.24 \pm 0.02$ |
| Trenberth | 1981 | $5.137 \pm 0.0002$ |
| Trenberth & Guillemot | 1994 | 5.1441 |
| Trenberth & Smith | 2005 | 5.1480 |
| Simpson (this analysis) | 2017 | 5.151 |

Unlike previous estimates that rely on surface pressure measurements or estimates, this work is an air-density based analysis that sums small density-volume contributions to $M_A$ from the earth's surface to the outer limits of the atmosphere. It uses the U. S. Standard Atmosphere, 1976 (USSA) model, along with two hypsographic curves – one being circa 1931 and the other from 2012. This approach provides estimates of the atmospheric mass in the various layers of the atmosphere and the mean and median surface elevations, in addition to providing insight into the role of the land contour.

Several different numerical techniques were applied for interpolation and integration. When feasible, two or more methods were used to maximize accuracy. Results are numerically precise to the number of digits reported in the tables. Final results are rounded to reflect uncertainty in the modeling assumptions and the physical data.

## U. S. Standard Atmosphere

Although dated, the USSA is probably the most commonly used atmospheric model even today. It has earned widespread acceptance as "an idealized, steady-state representation of the earth's atmosphere from the surface to 1,000 km …" It is a monumental work that involved 29 participating organizations, including National Oceanic and Atmospheric Administration (NOAA), National Aeronautics Administration (NASA), and the U.S. Air Force.

Most of the model as it relates to the mass of the atmosphere is based on the earlier 1962 version of the USSA model. However, little of significance has changed since then. The most striking change has been the increase in the $CO_2$ concentration, which has gone from 314 ppm (by volume), circa late 1950s, to 404 ppm in 2016. Assuming that the total kg moles of atmospheric gas has remained unchanged, but that some oxygen has been replaced with $CO_2$, it can be shown that atmospheric carbon has increased by $1.9 \times 10^{14}$ kg since the USSA work. While a large increase in carbon, it is too small to affect the value of the mass of the atmosphere reported here ($5.151 \times 10^{18}$ kg).

## Hypsographic Curves

Providing cumulative distributions of land elevations and ocean depths, hypsographic curves have been available since the late nineteenth century. They represent the fraction of the earth's total surface area that is land and is at or below the indicated elevation. *Wagner* [1895] prepared the first generally accepted curve. It was followed by *Kossinna*'s [1921] curve, which is widely referenced to this day even though he later updated it [*Kossinna* 1931, 1933]. Recently, *Eakins and Sharman* [2012] provided a modern version. Many variations of these curves are also in use.

These curves provide insight into how elevation predictions have changed over time. This is largely due to the difficulty in gathering the massive amounts of data needed to adequately define rugged land contours. In recent decades computers and satellite data have improved our understanding of topography. Reported mean land elevations by date are 700m (1895), 840m (1921), 871m (1931), and 797m (2012).

Here, attention is focused on the data used to construct two hypsographic curves, both of which are shown in Figure 1. The first curve was prepared from *Kossinna* [1931] data. He placed surface area incremental data into nine elevation intervals, the first being below sea level and last being above 5 km. The second curve is from the recent results of *Eakins and Sharman* [2012], which includes 6,700 data points from sea level to 8.333 km. It is drawn through all of the data points. The fraction of the earth's surface that is land is 0.292 and 0.2905 for the 1931 and 2012 estimates, respectively.

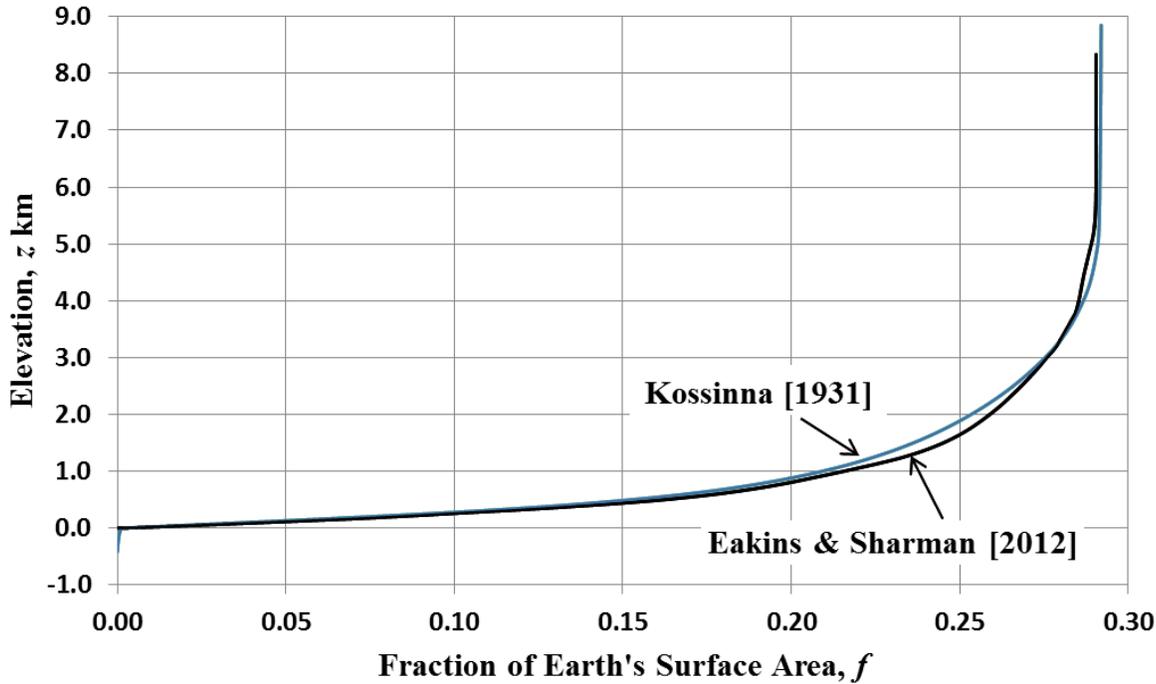

**Figure 1.** Hypsographic curves of Kossinna [1931] and Eakins & Sharman [2012].

To construct a mathematical curve from the Kossinna data it was necessary to interpolate between rather coarse elevation increments and to assign elevations to the end points. The end points selected were the lowest land elevation (the Dead Sea at -0.418 km) and the highest (Mt. Everest at 8.848 km). The curve had to (1) pass through the data points (knots), (2) have no maxima or minima, and (3) leave little or no evidence of the knots used to construct it. Regression fits and polynomial equations do not meet the first two criteria.

The modified "natural" cubic spline used here was the only technique found to meet all three criteria. The sea-level knot was the usual terminal condition ($d^2z/df^2 = 0$). At 5 km the slope ($dz/df$) was set equal to 400 km to minimize mathematical artifacts; however, results are insensitive to this value. Cubic equations were used for the areas from -0.418 km to sea level and from 5 km to 8.848 km. The second derivatives of these equations were set equal to zero at the lowest and highest elevations and the values and slopes were those of the spline at the 0 km and 5 km knots. The resultant curve is quite smooth; however, the data behind it is dated and the area contained in each elevation increment was reported to only three significant digits, which is equivalent to the nearest $10^5$ km$^2$.

Because the hypsographic curve due to *Eakins and Sharman* [2012] consists of 6,700 data points, cubic spline interpolation was not necessary when using that data set to estimate the mass of the atmosphere.

Although the curves appear similar, there are marked differences between 1 and 3 km. Elevation averages from the numerical analysis are summarized in Table 2. The earth-based mean of 231.3 m is very close to the value (231.74 m) recommended by *Trenberth and Smith* [2005]. Note that median values are more characteristic of the average elevation since they represent the elevation where half of the land area is higher and half is lower.

**Table 2.** Calculated elevation averages.

| Averages | Land (m / ft) | Earth (m / ft) |
|---|---|---|
| Kossinna [1931] | | |
|    Mean | 870.7 / 2,857 | 254.3 / 834 |
|    Median | 466 / 1,530 | |
| Amante & Eskins [2012] | | |
|    Mean | 796.1 / 2,612 | 231.3 / 759 |
|    Median | 420.5 / 1,380 | |

**Mass of the Atmosphere**

This analysis uses the USSA model, hypsographic curves, and numerical methods to calculate the mass of the atmosphere. First, it is appropriate to address the role of water vapor in the USSA model. Water vapor is a minor but significant component in the mass of the atmosphere. The average amount of precipitable water in the atmosphere is about 25 mm [*Trenberth and Guillemot*, 1994] with monthly variations of roughly 3 to 4 mm above and below that average. It is easily shown that 25 mm of water corresponds to 245 Pa, which is its average partial pressure in air. It is equally easy to show that 25 mm of water amounts to $1.27{\times}10^{16}$ kg of atmospheric mass, which is roughly 0.25 % of the total.

The USSA model is often referred to as being based on dry air. It is and it isn't. It is in the sense that the molecular weight is that of dry air and sea-level density, as well as altitude-related pressure and density changes, depend the molecular weight. However, the model is based on the standard sea-level pressure (101,325 Pa), which does include a water vapor contribution. There may be no way to adjust the USSA model to include water vapor short of reformulating the model itself. Even with such a modification, the result is likely to be a very small adjustment of the 0.25 % contribution. Here we assume that the USSA model properly accounts for the effect of water vapor.

A useful way to estimate atmospheric mass via density changes is to integrate the product of the USSA air density and thin spherical volume elements from sea level to outer space, and then subtract the air mass displaced by land, which is referred to here as the virtual mass. The total mass of the atmosphere is determined from Equation 1.

$$M_A = \int_0^\infty A\,\rho\,dz - \int_0^{f_L}\int_0^{z_L} A\,\rho\,dz\,df \tag{1}$$

where

$$A = A_e(1 + z/r_e)^2 \tag{2}$$

and    $A_e = 5.10065622{\times}10^{14}$ m$^2$, the surface area of the reference ellipsoid

       $r_e = 6{,}371{,}007.18$ m, the radius of a sphere with the area $A_e$

       $\rho = $ air density

       $z = $ geometric elevation

       $f_L = $ fraction of the earth's surface area that is land

       $z_L = $ elevation at the surface of the land above sea level, which is a function of $f$.

Note that even though the second integrand in Equation 1 seems to have no dependence on $f$, $f$ and $z$ are interdependent via the hypsographic curves.

Here the earth is considered to be a sphere with the area of the reference ellipsoid, as determined in WGS84 [2014]. Unless otherwise specified all elevations are considered to be geometric values.

Since the highest elevation above sea level is 8.848 km, all of the land-surface air densities are in Layer 1 of the USSA model, which includes altitudes below 11 km geopotential height. The elevation double integral in Equation 1 is the virtual mass ($M_V^*$); it can be evaluated from Equation 3. It is the atmospheric mass that would be necessary to maintain the standard pressure if all the earth were at sea level.

$$M_V^* = A_e \int_0^{f_L} \int_0^{z_L} (1 + z/r_e)^2 \, \rho_0 \left[ 1 + \frac{z \, L/T_0}{1 + z/r_2} \right]^{[-g_0 M/(R \, L) - 1]} dz \, df \qquad (3)$$

where $\rho_0$, $T_0$, and $g_0$ are respectively density (1.225 kg/m$^3$), temperature (288.15 K), and the USSA acceleration of gravity (9.80665 m/s$^2$) – all at sea level. In addition, $L$, $M$, $R$, and $r_2$ are the lapse rate (-0.0065 K/m), molecular weight (28.9644 kg/kg-mole), universal gas constant (8,314.32 Pa m$^3$/(kg-mol K)), earth radius used with the USSA formulation (6,356,766 m), respectively. For simplicity, we define $n = -g_0 M/(R \, L) = 5.25588$.

Both $z/r_e$ and $z/r_2$ are very small and could be neglected, which would allow Equation 3 to be integrated analytically. Instead, the equation was integrated numerically for several different elevations $z_L$ and a regression was performed. This procedure yielded the following equation:

$$M_V^* = \frac{A_e \rho_0 T_0}{nL} \int_0^{f_L} \left[ 1 + 1.0185 \frac{z_L}{r_e} + 61.51 \left( \frac{z_L}{r_e} \right)^2 \right] \left[ \left( 1 + \frac{z_L L}{T_0} \right)^n - 1 \right] df \qquad (4)$$

The maximum correction multiplier is 1.0015, obtained from the first bracketed term for $z_L = 8.848$ km. It is not feasible to calculate meaningful error limits for this type of estimate because of uncertainty associated with the many variables that are used to obtain the final result. These variables include but are not limited to the elevation distribution, the reference ellipsoid representing sea level, and the 1976 U. S. Standard Atmosphere formulation. Equation 4 was numerically integrated using the hypsographic data, resulting in $M_V^* = 1.31799 \times 10^{17}$ kg and $1.44329 \times 10^{17}$ kg for the *Eakins and Sharman* [2012] and *Kossinna* [1931] curves, respectively.

The first integral in Equation 1 was evaluated using extended trapezoidal and Simpson's rule integration, both with 1 m increments. In addition, 16 point Gaussian quadrature was applied to the first seven layers. Mass above 86 km was included but could have been neglected. The upper bounds on the layers are 11, 20, 32, 47, 51, 71, 86, and 1,000 km. Up to 86 km geometric height all calculations use geopotential height. The molecular weight correction in Table 8 of the USSA model was implemented with natural cubic spline interpolation from 80 to 86 km. Mass above 86 km is from the computer code of *Pietrobon* [1999] using extended Simpson's and trapezoidal rule integration. The best estimate of the total mass above sea level (i.e., the first integral in Equation 1) is $5.282300 \times 10^{18}$ kg.

A breakdown of the total mass of the atmosphere is given in Table 3. Even though the air mass in Layer 1 is reduced because of the land contours, it still contains more then 3/4 of the total mass.

**Table 3.** Total atmospheric mass by layer.

| Layer | Name | Maximum Height (km) | | Mass (kg) | |
|---|---|---|---|---|---|
| | | Geopotential | Geometric | Kossinna | Eakins & Sharman |
| 1 | Troposphere | 11 | 11.019 | 3.954367E+18 | 3.966897E+18 |
| 2 | Stratosphere | 20 | 20.063 | 8.964512E+17 | 8.964512E+17 |
| 3 | Stratosphere | 32 | 32.162 | 2.414449E+17 | 2.414449E+17 |
| 4 | Stratosphere | 47 | 47.350 | 3.983928E+16 | 3.983928E+16 |
| 5 | Stratosphere | 51 | 51.412 | 2.322310E+15 | 2.322310E+15 |
| 6 | Mesosphere | 71 | 71.802 | 3.335450E+15 | 3.335450E+15 |
| 7 | Mesosphere | 84.852 | 86.000 | 1.908676E+14 | 1.908676E+14 |
| 8 | Thermosphere | | 1000.000 | 2.056000E+13 | 2.056000E+13 |
| | | | Total | 5.137972E+18 | 5.150502E+18 |

The best estimate of the mass of the atmosphere from this analysis is $5.151 \times 10^{18}$ kg. It is in excellent agreement with the most recent estimate of *Trenberth and Smith* [2005] ($5.1480 \times 10^{18}$ kg). The lower estimate from analysis of the *Kossinna* [1931] data is consistent with earlier results from *Verniani* [1965] and *Trenberth* [1981].

Note that the first integral in Equation 1 represents the density-based atmospheric mass if all the earth were at sea level. It should equal the common estimate based on sea-level pressure, ellipsoid area, and average gravitational acceleration. Dividing this pressure-based estimate by the first integral in Equation 1, which is based on the USSA model, we define the constant $C$.

$$C = \frac{P_0 \, A_e / g_e}{\int_0^\infty A \, \rho \, dz} \tag{5}$$

With the previously determined value for the integral and $P_0 = 101,325$ Pa and $g_e = 9.7976432222$ m/s$^2$ (the global-average sea-level acceleration of gravity per WGS84 [2014]), $C = 0.99861$, which is remarkably close to unity, thereby validating the use of the USSA model. This constant can also be used to transform Equation 1 into the hybrid estimate of the atmospheric mass given in Equation 6.

$$M_{a,hybrid} = P_0 \, A_e / g_e - C \, M_V^* \tag{6}$$

This equation eliminates much of the uncertainty associated with use of the USSA model. With $M_V^* = 1.31799 \times 10^{17}$ kg from the *Eakins and Sharman* [2012] contour data, $M_{a,hybrid} = 5.143 \times 10^{18}$ kg. Only the virtual mass ($M_V^*$) is retained, which is only 2.6 % of the mass of the atmosphere.

**Acknowledgments**

We are grateful to Adolfo Figueroa-Viñas for reviewing the manuscript and contributing helpful suggestions.